\documentstyle[12pt]{article}
\begin{document}
\begin{titlepage}
\vspace*{-1.5cm}
\begin{center}
%
\hfill IC/95/25
%
\\[1ex]  \hfill February, 1995

%
\vspace{5ex}
{\Large \bf $(B+L)$-conserving Nucleon Decays
\\[1ex] in Supersymmetric Models}

%
\vspace{4ex} {\bf  Francesco Vissani }

%
{\it
\vspace{3ex} International Center for Theoretical Physics, ICTP
\\[0ex] Strada Costiera 11, I-34100 Trieste, Italy
\\[1ex] and
\\[1ex] Istituto Nazionale di Fisica Nucleare, INFN
\\[0ex] Sezione di Trieste, c/o SISSA
\\[0ex] Via Beirut 2-4, I-34013 Trieste, Italy
}
\vskip1truecm
\end{center}

%
\begin{quotation}
The presence of the $(B+L)$-conserving
decay modes
$n \to K^+ e^-,$
$n \to K^+ \mu^-,$
$p \to K^+ e^- \pi^+$ and
$p \to K^+ \mu^- \pi^+$
is shown to be a characteristic feature
of a class of models
with explicit breaking
of $R$-parity.
These modes dominate
over the $(B-L)$-conserving ones
in certain regions of the parameter space;
the impact of this scenario
for nucleon decay search
at the Super-Kamiokande
is discussed.
\end{quotation}
\end{titlepage}
\vfill\eject
%
There is a theoretical prejudice in favor of
$(B-L)$-conservation in nucleon decay,
supported by the following argument.
The lowest dimension, gauge invariant operators
that give rise to nucleon decay
involve necessarily three quarks and one
{\em antilepton} field
\cite{Weinberg}.
Higher dimensional operators can violate the
$\Delta B=\Delta L$ selection rule:
this is the case of dimension seven operators that
involve three quarks, one {\em lepton} and
one Higgs-doublet
(or one covariant derivative) \cite{Weinberg,dim7}.
If the effective operators
are generated by baryon- and lepton-number
violating interactions at the scale $M_x,$
it is reasonable to expect that the
contribution of the latter to the
nucleon decay rate is suppressed by powers
of $v/M_x$ at best, where $v$ is the scale
of $SU(2)_L$-symmetry breaking.

Under special conditions the previous argument can be evaded,
and $(B-L)$-violating decays can reach observable levels:
1) {\em If dimension six operators are not generated.}
This happens for instance if the four-fermions operators
responsible of nucleon decay are induced only after
$SU(2)_L$-breaking.
The four-fermions operators
generated  by the exchange of a scalar 10-plet
in an extension of minimal $SU(5)$ \cite{Wilczek-Zee}
fall in this category.
In a different scenario
the leading operators for nucleon decay
entail six fermions.
This has been explicitly realized in a
model based on the Pati-Salam
Unification group \cite{Pati-Salam-Sarkar}.
2) {\em If $v/M_x$ is not small.}
The present Letter is devoted to the exploration
of this possibility
in supersymmetric models.
In this context $M_x$ will be of the order of the mass of
scalar quarks, that we assume in the TeV range.

Let us assume the minimal supersymmetric matter content:
one superfield for each ordinary quark and lepton field
(for instance the $SU(2)_L$-doublet superfield $Q$ corresponds
to the the $q_L$ quark doublet, and
contains the $\tilde q_L$ scalar
quark).
Gauge invariance admits the following
terms in the superpotential \cite{lambda-terms}
($R$-parity breaking terms):
\begin{equation}
\lambda_{ijk}\cdot L_i L_j E^c_k+
\lambda_{ijk}'\cdot  D^{c,\alpha}_i L_j Q_k^\alpha+
\lambda_{ijk}''\cdot \epsilon_{\alpha\beta\gamma}
D_i^{c,\alpha}D_j^{c,\beta}U_k^{c,\gamma},
\label{lambda terms}
\end{equation}
where $\alpha,\beta,\gamma$ are color indices,
$i,j,k$ are family indices and $SU(2)_L$-doublets
are contracted by $i\tau_2$.
It is  convenient to consider these couplings in the basis
in which $u_{R,i},$ $d_{R,i},$ $e_{R,i},$
$u_{L,i}$ and $e_{L,i}$ are mass eigenstates.
Notice that the couplings $\lambda$ and $\lambda''$ are
antisymmetric in the first two indices.

The exchange of $\tilde d_{R,i}$
squarks, in the hypothesis of mass degeneracy,
gives rise to the following $(B-L)$-conserving operators
\cite{Hincliffe-Nir}:
\begin{equation}
\frac{2\ \lambda_{il1}''{}^*
\lambda_{ijk}'}{m_{{\tilde d}_{R}}^2}\cdot
\epsilon_{\alpha\beta\gamma}
(d_{R,l}^{\alpha} u_{R}^{\beta})
(e_{L,j} u_{L,k}^\gamma-V^{\rm\scriptscriptstyle CKM}_{kn}\
\nu_{L,j} d_{L,n}^\gamma),
\label{induced-b-l}
\end{equation}
where $V^{\rm\scriptscriptstyle CKM}$ is the
Cabibbo-Kobayashi-Maskawa matrix.
The operators relevant for nucleon decay involve:
$e,\ \mu$ or $\nu_1,\ \nu_2,\ \nu_3$ lepton fields;
$u$  or $d,\ s$ quark fields (but no more than
one $s$ field).
The required suppression of the decay
rate is due to
the smallness of the $\lambda$-couplings.

New kinds of four fermions operators are induced
when the mixing of squarks contained in the $Q$
and in $D^c$ (or $U^c$) supermultiplets
is taken into account, since
the $B$-violating and $L$-violating interaction terms
coming from (\ref{lambda terms})
can be also connected
by the $\langle\tilde q_R\ \tilde q_L^*\rangle$ propagator.
The induced operators are of the form
$(q q)(\bar l q),$ and therefore conserve $B+L;$
they will be discussed in the following,
after recalling the relevant informations
on the squark mass matrix.

The $6\times 6$ matrix for up- or down-type squarks
can be written as:
\begin{equation}
M_{\tilde q}^2=
\left(
\begin{array}{cc}
m^2_{\tilde q,{\scriptscriptstyle LL}}&
m^2_{\tilde q,{\scriptscriptstyle LR}}\\
m^{2*}_{\tilde q,{\scriptscriptstyle LR}}&
m^2_{\tilde q,{\scriptscriptstyle RR}}
\end{array}
\right).
\label{squark matrix}
\end{equation}
To avoid constraints from nonobservation of
flavor-changing neutral currents it is usually
posited that the off-diagonal elements
are small with respect to the diagonal ones.
But this condition is not expected to hold
for the stop and sbottom left-right mixing terms, since they
are proportional to the masses of the heaviest quarks:
\begin{equation}
\begin{array}{rcl}
m^2_{\tilde t,{\scriptscriptstyle LR}}&=&
m_t (A_t+\mu\ {\rm cot}\beta)\\
m^2_{\tilde b,{\scriptscriptstyle LR}}&=&
m_b (A_b+\mu\ {\rm tan}\beta)
\end{array}
\label{LR-tb}
\end{equation}
[$A_t$ and $A_b$ are the parameters of the Higgs-squark-squark,
soft-supersymme\-try-breaking interactions;
$\mu$ is the supersymmetry-conserving Higgs mass parameter;
$\tan\!\beta$ the ratio of vacuum expectation values
$\langle H_u\rangle / \langle H_d\rangle$].
For this reason it is important to include
the effect of the stop and of the sbottom
left-right mixing terms
in the discussion.

In the limit of small momenta
the propagator $\langle\tilde q_R\
\tilde q_L^*\rangle$ ($q=t,\ b$)
reduces to
\begin{equation}
\frac{-i}{{\cal M}^2_{\tilde q}}\equiv
\frac{
i\ m_{\tilde q,{\scriptscriptstyle LR}}^2
}{
m_{\tilde q,{\scriptscriptstyle LL}}^2\
m_{\tilde q,{\scriptscriptstyle RR}}^2
-m_{\tilde q,{\scriptscriptstyle LR}}^4},
\label{kinematic}
\end{equation}
The factor $1/{{\cal M}^2_{\tilde q}}$
can be compared with
$1/m_{{\tilde d}_{R}}^2$
in eq.\ (\ref{induced-b-l}).
The contribution of (\ref{kinematic}) is enhanced by
a large left-right mixing, as well as by
a light $\tilde q$ squark,
since the denominator of the
right hand side is the determinant of the stop or of the sbottom
mass matrix \cite{cw}.

The $\tilde b_L-\tilde b_R$ exchange
gives rise to the following operators:
\begin{equation}
\frac{(2\ \lambda_{3l1}''\
\lambda_{nj3}')^*}{{\cal M}^2_{\tilde b}}
\cdot
\epsilon_{\alpha\beta\gamma}
(d^\alpha_{R,l} u_R^\beta)
(\bar \nu_{R,j} d^\gamma_{R,n}),
\label{induced-b+-l}
\end{equation}
for which the remarks done for (\ref{induced-b-l}) also apply.
However this kind of $(B-L)$-violations
cannot be experimentally distinguished from
a $(B-L)$-conserving signal,
since the neutrinos are undetectable.

The case of virtual stop exchange is more interesting.
We get the following contributions in the
effective lagrangian:
\begin{equation}
-\frac{(2\ \lambda_{123}''\
\lambda_{1j3}')^*}{{\cal M}^2_{\tilde t}}
\cdot
\epsilon_{\alpha\beta\gamma}(d^\alpha_R s^\beta_R)
(\bar e_{R,j} d^\gamma_R),
\label{induced-b+l}
\end{equation}
where
$j=1$ corresponds to the electron channel,
and $j=2$ to the muon.

The $\lambda$-couplings appearing
in the previous expression
differ from those
in (\ref{induced-b-l}) and in (\ref{induced-b+-l}).
Therefore it is interesting
to speculate upon the scenario in which
the decay modes induced by (\ref{induced-b+l}) are
dominating ones.
These decay modes are
\begin{equation}
n\to K^+ l^-
\label{2-bodies}
\end{equation}
and
\begin{equation}
\begin{array}{r}
n\to K^+ l^- \pi^0\\
n\to K^0 l^- \pi^+\\
p\to K^+ l^- \pi^+,
\end{array}
\label{3-bodies}
\end{equation}
where $l=e,\ \mu.$

The main contribution to the nucleon decay rate is
expected to come from the two-body decays: in fact
estimating $\Gamma\sim |\lambda'\lambda''|^2\
\Delta m^5/{\cal M}^4,$
where $\Delta m$ is the energy release, we obtain
$\Gamma$(three-body)/$\Gamma$(two-body)$\sim 15$ \%
for the electron channels, $\sim 7$ \% for the muon channels.
The proton decay rate may be an order of magnitude slower
than the neutron three-body decay rates,
since it proceeds via
the quark-decay mechanism, whereas the neutron decays
occur via the more efficient
two-quark fusion mechanism (this
argument has been proposed in Ref.\ \cite{JY}).

The mode in eq.\ (\ref{2-bodies}) and the
proton  decay mode in eq.\ (\ref{3-bodies})
have already been searched by various experimental groups
\cite{PDG-Frejus}, and the lower bounds on the lifetime
translate
in the upper bound
\begin{equation}
\left|
\frac{\lambda''_{123}\ \lambda'_{1j3}}{{\cal M}^2_{\tilde t}}
\right|
{\ \raisebox{-.4ex}{\rlap{$\sim$}}
\raisebox{.4ex}{$<$}\ } 10^{-30}
{\rm GeV}^{-2},
\label{upper-bound}
\end{equation}
where $j=1,2.$
This bound is saturated by
$|\lambda_{123}'' \lambda_{1j3}'|\sim 10^{-26},$ if we
assume  ${\cal M}_{\tilde t}\sim 1$ TeV.

Future search at the Super-Kamiokande \cite{sk}
may improve the bound (\ref{upper-bound}) by
more than an order of magnitude.
It is important to remark
that the decay $n\to K^+ l^-$ provides a quite clear signal
in water \v{C}erenkov detectors:
\begin{equation}
{}^{16}O\to {}^{15}O + \gamma(6.2\ {\rm MeV}) + \mu + l,
\label{exp-sign}
\end{equation}
where $l$ is monochromatic, $\mu$ results from kaon decay
and $\gamma$ from the transition of the excited nucleus
to the ground state.
Since  \v{C}erenkov-light detectors cannot distinguish
the charges of particles, the signal (\ref{exp-sign})
may be equally well interpreted as the result of the
$(B-L)$-conserving $n\to K^- l^+$ decay.
But this two cases are different if one
takes into account also other channels.
In fact the lowest dimension operator that induces
$n\to K^- l^+$ is of the form
$d d u u \bar s l.$
This implies that
the channels $p\to \bar K^0 l^+$ and
$p\to \bar K^0 l^+ \pi^0$ are opened as well.
Moreover, if six-fermions operators dominate,
there is no a-priori reason to expect the suppression of
the strangeness-conserving nucleon decays.
Instead no other two-body channel is opened
in the $(B+L)$-conserving scenario
for nucleon decay
proposed in the present Letter.

We conclude that the
stop-mediated nucleon decays
are well characterized at the Super-Kamiokande.
This result motivates the experimental search
of the kind of signals described
above and the theoretical search
of a natural framework for
small $R$-parity violating couplings.
%

%
I would like to thank
A. Salam for hospitality at the Centre;
A. Yu. Smirnov for invaluable help;
G. Senjanovi\'c, B. Brahmachari, S. Bertolini,
F. Hussain, and G. Thompson
for interesting discussions.

%


\begin{thebibliography}{99}
\bibitem{Weinberg}
S. Weinberg, {\it Phys. Rev. Lett.} {\bf 43}, 1566 (1979);
F. Wilczek and A. Zee,
{\it Phys. Rev. Lett.} {\bf 43}, 1571 (1979).

\bibitem{dim7}
S. Weinberg, {\it Phys. Rev.} {\bf D 22}, 1694 (1980);
H. A. Weldon and A. Zee,
{\it Nucl. Phys.} {\bf B 173}, 269 (1980).

\bibitem{Wilczek-Zee}
F. Wilczek and A. Zee, {\it Phys. Lett.} {\bf B 88}, 311 (1979).

\bibitem{Pati-Salam-Sarkar}
J. Pati, A. Salam and U. Sarkar,
{\it Phys. Lett.} {\bf B 133}, 330 (1983).

\bibitem{lambda-terms}
S. Weinberg, {\it Phys. Rev.} {\bf D 26}, 287 (1982);
N. Sakai and T. Yanagida,
{\it Phys. Lett.} {\bf B 197}, 533 (1982).

\bibitem{Hincliffe-Nir}
I. Hincliffe and T. Kaeding,
{\it Phys. Rev.} {\bf D 47}, 279 (1993);
Y. Nir and V. Ben-Hamo,
{\it Phys. Lett.} {\bf B 339}, 77 (1994).


\bibitem{cw}
For low values of $\tan\!\beta,$ in the approximation of
M. Carena, M. Olechowski, S. Pokorski and C.E.M. Wagner,
{\it Nucl. Phys.} {\bf B 419}, 213 (1994),
we find  that $|1/{\cal M}^2_{\tilde t}|$ is larger than
$1/m_{{\tilde d}_{R}}^2$ for gluino masses
${\ \raisebox{-.4ex}{\rlap{$\sim$}}
\raisebox{.4ex}{$<$}\ } 500$ GeV
(neglecting the $\mu \cot\beta$ contribution to
$m_{\tilde t,{\scriptscriptstyle LR}}^2$).
For large values,
$\tan\!\beta{\ \raisebox{-.4ex}{\rlap{$\sim$}}
\raisebox{.4ex}{$>$}\ } m_t/m_b,$ also
$|1/{\cal M}^2_{\tilde b}|$ can be of the same
order of magnitude.


\bibitem{JY}
C. Jarlskog and F.J. Yndur\'{a}in,
{\it Nucl. Phys.} {\bf B 149}, 29 (1979).

\bibitem{PDG-Frejus}
Fr\'ejus Collab., Ch. Berger et al.,
{\it Phys. Lett.} {\bf B 269}, 227 (1991);
PDG, {\it Phys. Rev.} {\bf D 50}, 1173 (1994).

\bibitem{sk}
Y. Totsuka,
Report no. ICRR-277-90-20, 1990 (unpublished).

\end{thebibliography}
\end{document}